\documentclass[
]{ceurart}

\usepackage{listings}
\usepackage{amsmath,graphicx}
\usepackage{tabularx}
\usepackage{pifont}

\begin{document}

\copyrightyear{2025}
\copyrightclause{Copyright for this paper by its authors. Use permitted under Creative Commons License Attribution 4.0 International (CC BY 4.0).}

\conference{SeMatS 2025: The 2nd International Workshop on Semantic Materials Science co-located with the 24th International Semantic Web Conference (ISWC 2025), November 3rd, 2025, Nara, Japan}

\title{Semantic Representation of Processes with Ontology Design Patterns}

\author[1,2]{Ebrahim Norouzi}[%
orcid=0000-0002-2691-6995,
email=Ebrahim.Norouzi@fiz-Karlsruhe.de,
]
\cormark[1]
\author[1,3]{Sven Hertling}[%
orcid=0000-0003-0333-5888,
email=sven.hertling@fiz-karlsruhe.de,
]
\author[1]{Jörg Waitelonis}[%
orcid=0000-0001-7192-7143,
email=joerg.waitelonis@fiz-karlsruhe.de,
]
\author[1,2]{Harald Sack}[%
orcid=0000-0001-7069-9804,
email=Harald.Sack@fiz-Karlsruhe.de,
]

\address[1]{FIZ Karlsruhe -- Leibniz Institute for Information Infrastructure, Hermann-von-Helmholtz-Platz 1, 76344 Eggenstein-Leopoldshafen, Germany}
\address[2]{Karlsruhe Institute of Technology, Institute of Applied Informatics and Formal Description Methods, Kaiserstr. 89, 76133 Karlsruhe}
\address[3]{University of Mannheim, Data and Web Science Group, School of Business Informatics and Mathematics B6 26, 68159 Mannheim}

\cortext[1]{Corresponding author.}

\begin{abstract}
The representation of workflows and processes is essential in materials science engineering, where experimental and computational reproducibility depend on structured and semantically coherent process models. Although numerous ontologies have been developed for process modeling, they are often complex and challenging to reuse. Ontology Design Patterns (ODPs) offer modular and reusable modeling solutions to recurring problems; however, these patterns are frequently neither explicitly published nor documented in a manner accessible to domain experts. This study surveys ontologies relevant to scientific workflows and engineering process modeling and identifies implicit design patterns embedded within their structures. We evaluate the capacity of these ontologies to fulfill key requirements for process representation in materials science. Furthermore, we propose a baseline method for the automatic extraction of design patterns from existing ontologies and assess the approach against curated ground truth patterns. All resources associated with this work—including the extracted patterns and the extraction workflow—are made openly available in a public GitHub repository.\footnote{\url{https://github.com/ISE-FIZKarlsruhe/odps4mse}}.
\end{abstract}

\begin{keywords}
  Ontology \sep
  Ontology design patterns \sep
  Materials science \sep
  Scientific workflow \sep
  Process modeling
\end{keywords}

\maketitle

\section{Introduction}
\label{sec:Introduction}

Process modeling is a broad field that plays a critical role across numerous scientific and engineering domains, particularly in Materials Science and Engineering (MSE). For decades, the development of new materials has relied on understanding the intricate correlation between processing methods and resulting material properties \cite{olson_material_process}. The well-established paradigm of processing–structure–properties–performance encapsulates the fundamental principles of MSE: material performance is governed by its properties, which are determined by its structure, and the structure is ultimately shaped by the applied processing route \cite{ashby1993materials, callister2020callister}.  

Accurately modeling processing steps is therefore essential for representing the dynamic transformations that occur during the material's life cycle. Several large-scale initiatives\footnote{\url{https://www.mgi.gov/}} \footnote{\url{https://www.plattform-i40.de}} \footnote{\url{https://www.materialdigital.de/}} \footnote{\url{https://nfdi-matwerk.de/}} \cite{bayerlein2024pmd} aim to formalize materials data and workflows through taxonomies, ontologies, and knowledge graphs, encompassing the entire life cycle from synthesis to performance evaluation. A key challenge, however, lies in capturing the dynamic events during which materials transition between states due to processing interventions\cite{alam2021ontology_pmd}.  

Multiple ontologies have been developed for process modeling in MSE, yet issues with interoperability and reuse persist. A major gap is the limited adoption of Ontology Design Patterns (ODPs) \cite{norouzi2024landscape}, which provide reusable template solutions for recurring modeling problems \cite{gangemi2009ontologydesignpatter}. In many cases, ontology developers or domain experts embed these patterns unintentionally, failing to explicitly publish them as reusable modules \cite{blomqvist2015considerations}. Identifying and extracting these embedded patterns could significantly enhance ontology reuse and provide domain experts with modular building blocks for representing workflows and processes without requiring them to navigate entire ontologies.

In this work, we focus on scientific workflow modeling and the extraction of ontology design patterns relevant to process representation in MSE. We achieve this by surveying existing ontologies, identifying patterns that align with domain-specific requirements, and proposing a baseline method for automatically extracting ODPs using semantic similarity techniques.

The remainder of this paper is structured as follows. We begin in Section~\ref{sec:relatedwork} with a review of related ontologies and existing ontology design patterns for process modeling. Section~\ref{sec:methodology} details our methodology for automatically extracting and evaluating these patterns. In Section~\ref{sec:results}, we discuss the patterns we extracted based on our defined requirements and present an evaluation of the results. Finally, Section~\ref{sec:Conclusion} provides our conclusion and suggests directions for future research.

\section{Related Work}
\label{sec:relatedwork}

Numerous ontologies have been developed to model processes, workflows, and experimental activities. However, their relevance to Materials Science and Engineering (MSE) depends on their domain focus and the granularity of their process semantics.

Several general-purpose ontologies have been created to represent scientific experimentation. For instance, the EXPO ontology\footnote{\url{http://expo.sourceforge.net/}} was designed to formalize experimental design, methodology, and result representation across various disciplines \cite{expo}. Although comprehensive, it lacks the specificity required to model the technical details of MSE processes. In the life sciences, ontologies like EXACT2\footnote{\url{https://bmcbiomed.biomedcentral.com/articles/10.1186/1471-2105-15-S14-S5}} \cite{soldatova2014exact2} and SMART Protocols\footnote{\url{https://smartprotocols.org/}} \cite{sp_ontology} were developed to capture biomedical and biological experimental protocols. These ontologies are domain-specific and not directly applicable to engineering workflows considered in this study.

In contrast, process ontologies in the engineering domain often focus on industrial or chemical workflows. OntoCAPE\footnote{\url{https://www.avt.rwth-aachen.de/Ontocape}} \cite{morbach2007ontocape}, for example, is a large-scale ontology for computer-aided process engineering. It models chemical plant design and operation, making it more suitable for the chemical industry than for experimental workflows common in materials science. The Procedural Knowledge Ontology (PKO)\footnote{\url{https://w3id.org/pko}} \cite{carriero2025pko} was developed to manage and reuse procedural knowledge, capturing procedures as sequences of executable steps. However, its focus lies on execution tracking and procedural documentation, not on modeling the structure of processes. Similarly, the BPMN-Based Ontology (BBO)\footnote{\url{https://www.irit.fr/recherches/MELODI/ontologies/BBO}} \cite{annane2019bbo} represents business-oriented process models and organizational workflows using the BPMN standard, which differs in scope from the scientific workflows relevant to this study.
 
Recently, a number of ontologies have been developed specifically to support workflow and process modeling in MSE \cite{norouzi2024landscape}. PMDcore\footnote{\url{https://w3id.org/pmd/co/2.0.8}} \cite{bayerlein2024pmd} describes material processing activities and ensures semantic interoperability in research data. The General Process Ontology\footnote{\url{https://gpo.ontology.link\#}} \cite{GPO} generalizes engineering process structures, enabling the composition of complex processes. The Workflows in Linked Data (WILD)\footnote{\url{http://purl.org/wild/vocab\#}} \cite{WILD} ontology supports the modeling of executable workflows in RDF, while P-PLAN\footnote{\url{http://purl.org/net/p-plan\#}} \cite{p-plan} enables the representation of scientific planning activities. Similarly, Metadata4Ing\footnote{\url{http://w3id.org/nfdi4ing/metadata4ing\#}} \cite{M4I} supports the documentation of data-generating processes in scientific experiments. The BWMD ontology~\cite{bwmd} was developed to support the semantic representation of material-intensive process chains. While the ontology offers a rich process modeling structure, we do not include it in our analysis due to its limited use of annotation properties, which makes it challenging to extract patterns and understand the modeling intent.

Foundational ontologies have also been used to construct reusable ontology design patterns (ODPs) for process modeling. A prominent example is the Basic Formal Ontology (BFO)\footnote{\url{https://basic-formal-ontology.org/}}, which serves as the top-level ontology for many process-centric models in the Open Biological and Biomedical Ontology (OBO) Foundry\cite{bfo2015}. A comprehensive review by Dooley et al.~\cite{dooley2024food} systematically examined BFO-based process models and their associated patterns across the OBO ecosystem. Since that study provides an in-depth analysis of BFO-derived patterns, these are not re-evaluated in this work. Another example is NFDICore\footnote{\url{https://ise-fizkarlsruhe.github.io/nfdicore/}}, which builds on BFO and defines patterns for data handling, contribution, and publishing processes \cite{bruns2024nfdicore}, with an emphasis on agent–information interactions\footnote{\url{https://ise-fizkarlsruhe.github.io/nfdicore/docs/patterns/\#processes}}. Although NFDIcore is well-suited for representing metadata about research resources within the National Research Data Infrastructure (NFDI) programme\footnote{\url{https://www.nfdi.de}}, its focus does not fully align with the process modeling needs of MSE, which emphasize material transformations and experimental workflows. Alternatively, the PlansLite ontology\footnote{\url{http://www.ontologydesignpatterns.org/ont/dul/PlansLite.owl}}, based on the DOLCE foundational ontology \cite{borgo2022dolce}, captures planning constructs through a task-based structure. While suitable for modeling general workflow logic, its notion of “process” is defined as a placeholder for evolving events not strictly tied to agents, tasks, or plans. As a result, it lacks the formalization needed to represent material-centered processes and transformations.

Beyond foundational ontologies, various domain-independent ontology design patterns (ODPs) have been proposed to address recurring challenges in process modeling. These include patterns such as Algorithm Implementation Execution\footnote{\url{https://odpa.github.io/patterns-repository/AlgorithmImplementationExecution/AlgorithmImplementationExecution}} \cite{lawrynowicz2017algorithm}, Material Transformation\footnote{\url{https://odpa.github.io/patterns-repository/Material_Transformation/Material_Transformation}} \cite{vardeman2017ontology_transformation}, and Reactive Processes\footnote{\url{https://odpa.github.io/patterns-repository/Reactor_pattern/Reactor_pattern}} \cite{solanki2012reactive}. Additional patterns like Sequence\footnote{\url{https://odpa.github.io/patterns-repository/Sequence/Sequence}}, Transition\footnote{\url{https://odpa.github.io/patterns-repository/Transition/Transition}} \cite{Carriero_2021_sequence}, Task Execution\footnote{\url{https://odpa.github.io/patterns-repository/TaskExecution/TaskExecution}}, Activity Reasoning\footnote{\url{https://odpa.github.io/patterns-repository/An_Ontology_Design_Pattern_for_Activity_Reasoning/An_Ontology_Design_Pattern_for_Activity_Reasoning}}, and Activity Specification\footnote{\url{https://odpa.github.io/patterns-repository/ActivitySpecification/ActivitySpecification}} \cite{katsumi2017_activity_defining} are perticularly useful for capturing the semantics of tasks, actions, and events. These patterns provide modular constructs for representing execution order and dependencies between tasks. In manufacturing domains, the VDI 3682 ODP\footnote{\url{https://github.com/hsu-aut/IndustrialStandard-ODP-VDI3682}} \cite{9097408_VDI_3682} and standards such as ISO 22400-2\footnote{\url{https://github.com/hsu-aut/IndustrialStandard-ODP-ISO22400-2}}, ISO 10303-44 (STEP)\footnote{\url{https://www.hsu-hh.de/aut/forschung/forschungsthemen/ontology-engineering-for-collaborative-embedded-systems/ontology-design-patterns-for-the-manufacturing-domain/iso-10303-44-product-structures}}, and DIN EN 62264-2\footnote{\url{https://github.com/hsu-aut/IndustrialStandard-ODP-DINEN62264-2}} address the need to unify and structure heterogeneous data environments across production systems. 

Despite their general applicability, these patterns have seen limited reuse in ontology development within the MSE community. One contributing factor is that many of these patterns are indeed too generic to fully capture the process modeling requirements of MSE contexts. However, it is essential to recognize that the adoption of semantic web technologies in MSE, including domain-specific ontologies and ontology design patterns, is still in its early stages. Consequently, even resources specifically designed for MSE have not yet achieved widespread uptake, often due to constraints such as a lack of tooling support and the steep learning curve for non-semantic web experts. In this sense, making ontology design patterns more accessible—both conceptually and technically—can help improve the adaptability and reuse of existing semantic assets in MSE.

In this work, we focus on the subset of ontologies and patterns most relevant to scientific workflow and process modeling in MSE. Specifically, we analyze PMDcore, the EMMO General Process Ontology, WILD, P-PLAN, and Metadata4Ing to identify reusable ontology design patterns that align with the defined process modeling requirements in materials science and engineering.

\section{Methodology}
\label{sec:methodology}

The methodology for this study was designed to identify and evaluate ontology design patterns (ODPs) for process modeling in Materials Science and Engineering (MSE). Guided by a review from domain experts, our approach focuses on capturing reusable patterns that address key requirements for representing scientific workflows and engineering processes. We began by selecting candidate ontologies based on their relevance to scientific workflows and material processing \cite{norouzi2024landscape}. From these, we extracted ODPs by isolating coherent modules that align with our defined requirements.

\subsection{Requirements Definition}  
The first step in our methodology involved defining the core requirements for ontology-based process modeling in MSE. These requirements were derived from an analysis of existing process ontologies and adapted to the specific needs of the MSE domain. The following three requirements guided the subsequent pattern extraction.  
\begin{itemize}
    \item {\bf Requirement 1: Process Structure}\\  
The ontology must describe the organization of processes, including their constituent steps, sub-processes, and execution order. In MSE, this corresponds to modeling transformations such as heat treatments, alloy mixing, thin-film deposition, and sequential multi-step processing stages that directly influence material microstructures.

\item {\bf Requirement 2: Data and Resources}\\  
The ontology must represent the flow of inputs and outputs, as well as the parameters that define each process step. For MSE, this specifically refers to capturing experimental parameters such as temperature, pressure, atmosphere conditions, measurement devices, and calibration data that are essential for reproducibility.

\item {\bf Requirement 3: Project Goals and Participant Roles}\\
The ontology must capture project goals, experimental stages, and their organizational context, including the agents involved and their specific roles within the project or process. In MSE, this relates to modeling multi-lab collaborations, research campaigns, project milestones, and the allocation of responsibilities (e.g., sample synthesis, microscopy, or simulation).
\end{itemize}

The extracted requirements and patterns were evaluated by nine domain experts within the Linked Open Data Working Group (LOD-WG)\footnote{\url{https://iuc12-nfdi-matwerk-ta-oms-7fd4826d9051b0dd93b21aa77d06d1d8c71c4.pages.rwth-aachen.de/docs/meetings/2025/}}. This group, a part of the \textit{National Research Data Infrastructure for Materials Science and Engineering (NFDI-MatWerk)}\cite{eberl_2021_5082837}\footnote{\url{https://nfdi-matwerk.de/}}, focuses on creating and reusing Linked Open Data across the MSE domain. Feedback from the LOD-WG was crucial for refining the extracted ODPs and validating their alignment with the domain-specific requirements of process modeling in MSE.

Ontology design patterns were first extracted and evaluated against our defined requirements. Following this, a baseline approach was developed to automatically identify these ODPs. The details of this method are presented in the subsequent section.

\subsection{Baseline Method for Automated ODP Extraction}
\label{sec:baseline}

\begin{figure}[h]
    \centering
    \includegraphics[width=\textwidth]{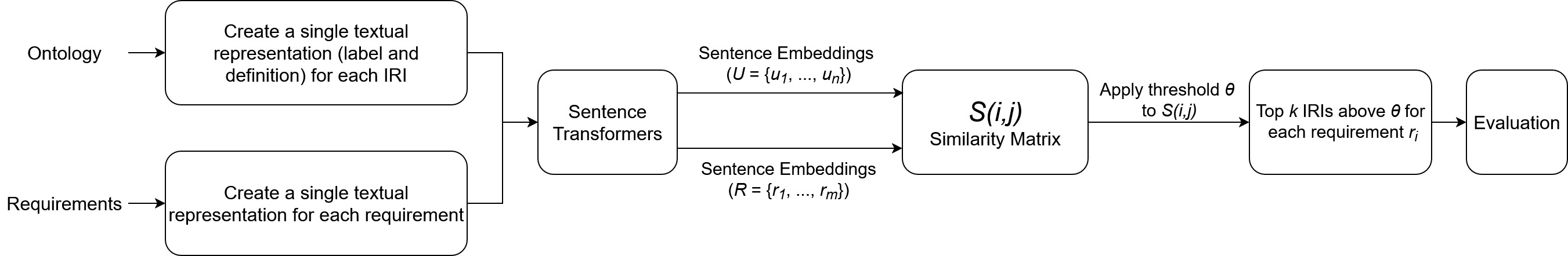}
    \caption{Baseline workflow for automatic extraction of ontology design patterns (ODPs).}
    \label{fig:odp-baseline-approach}
\end{figure}

Our methodology applies semantic similarity techniques to align ontology concepts with textual requirements, as illustrated in Figure~\ref{fig:odp-baseline-approach}. Each requirement is reformulated into natural language sentences to clearly express the intended representation.
For each ontology, we extract textual metadata from its classes and properties, specifically the values of \texttt{rdfs:label}, \texttt{skos:definition}, \texttt{rdfs:comment}, and related annotation properties. These are concatenated into unified textual descriptions for each IRI (Internationalized Resource Identifier).
We then embed both the requirements and ontology IRIs into a shared vector representation using sentence transformer models\cite{reimers-2019-sentence-bert}. A similarity matrix \( S(i, j) \) is computed between every requirement \( r_i \) and ontology concept \( u_j \), using cosine similarity. We apply a similarity threshold \( \theta \) to identify relevant ontology terms for each requirement.

The selected IRIs are evaluated against manually curated ground truth mappings using standard information retrieval metrics: precision, recall, and the $F_1$-score. These metrics are defined as follows:

Let \( C_i \) denote the set of ontology concepts retrieved for requirement \( r_i \), and \( C'_i \) the set of ground truth concepts for \( r_i \).
Precision is the proportion of correctly retrieved ontology IRIs out of all retrieved IRIs:
    \[
    \text{Precision} = \frac{|C_i \cap C'_i|}{|C_i|}
    \]
Recall is the proportion of correctly retrieved ontology IRIs relative to the total number of relevant IRIs:
    \[
    \text{Recall} = \frac{|C_i \cap C'_i|}{|C'_i|}
    \]
\( F_1 \)-score is the harmonic mean of precision and recall:
    \[
    F_1 = \frac{2 \cdot \text{Precision} \cdot \text{Recall}}{\text{Precision} + \text{Recall}}
    \]
To generate reusable ontology design patterns (ODPs) aligned with domain-specific requirements in materials science, a modular extraction strategy was applied using the ROBOT command-line framework.\footnote{\url{https://robot.obolibrary.org/extract}} This process transformed the identified IRIs into ontology modules, with ROBOT preserving the relevant structural and logical relations around the selected terms.

We evaluated four ROBOT-supported extraction methods: STAR, BOT, TOP, and subset. 
\begin{itemize}
\item \textbf{STAR} extracts a minimal module containing only the specified terms and those directly involved in logical entailments. It avoids a hierarchical structure, resulting in a compact module. 
\item \textbf{BOT} builds a module that includes seed terms and all their superclasses, preserving hierarchical context in a bottom-up manner. 
\item \textbf{TOP} includes the seed terms along with their subclasses, typically yielding larger modules. 
\item \textbf{Subset} materializes existential relations (e.g., \texttt{someValuesFrom}) and includes inferred axioms between seed terms. This method is ideal for extracting patterns that involve complex class expressions. 
\end{itemize}
We excluded the Minimum Information to Reference an External Ontology Term (\textit{MIREOT})~\cite{doi:10.3233/AO-2011-0087} method from our evaluation. Although useful for referencing external terms while minimizing dependencies, its lack of entailment preservation makes it unsuitable for logic-based pattern extraction. Each of the four extraction methods was also tested with varying levels of intermediate class inclusion using the  \texttt{all}, \texttt{minimal}, and \texttt{none} options.
\begin{itemize}
\item \texttt{all} includes all intermediate classes between seed terms and ontology roots. 
\item \texttt{minimal} retains only those intermediates with multiple siblings, preserving a minimal hierarchy.
\item \texttt{none} removes all intermediate classes unless they are explicitly referenced in logical axioms.
\end{itemize}
We selected the \texttt{STAR} method with \texttt{--intermediates none} option as our default configuration after an empirical evaluation. This setup generated semantically coherent and compact modules, preserving just enough logic to enable accurate pattern identification without extraneous context. For each requirement, we curated a list of seed terms and then extracted a separate ontology module accordingly.

\section{Results and Discussion}
\label{sec:results}

\subsection{Process Modeling Patterns}
\label{sec:results_patterns}

This section presents the manually derived design patterns, which were reviewed and evaluated by domain experts, and mapped to our defined requirements: Process ODP (Req.~1), Resource ODP (Req.~2), and Project ODP (Req.~3). In the accompanying figures, the ODPs are visually grouped using dashed red boundaries with corresponding labels.
While the above descriptions emphasize the manually derived structures, it is important to highlight how these patterns are subsequently integrated into the automated workflow (Section \ref{sec:baseline}. In practice, the extracted modules were directly reused for semantic similarity evaluation and ODP benchmarking (see Section \ref{sec:results_evaluation}). This ensures that the manually curated insights are not standalone but feed into the workflow’s automated pattern extraction and validation steps.

\paragraph{P-PLAN Ontology}  
P-PLAN supports both Process and Resource ODPs. Process structures are represented using \texttt{p-plan:Step}, \texttt{p-plan:Activity}, with their sequencing handled by \texttt{p-plan:isPrecededBy}. Inputs and outputs are modeled with \texttt{p-plan:Variable} and the properties \texttt{p-plan:hasInputVar}/\texttt{hasOutputVar}. This configuration provides a compact but expressive model for basic workflow representations. In this work, we focus on P-PLAN as it provides explicit constructs for modeling scientific planning activities and thus directly supports our defined requirements for process and resource ODPs. We did not include the core PROV ontology or other extensions such as ProvONE and D-PROV in our analysis, since our emphasis was on ontologies that are already specialized toward workflow representation. 
\begin{figure}[h]
    \centering
    \includegraphics[width=0.85\textwidth]{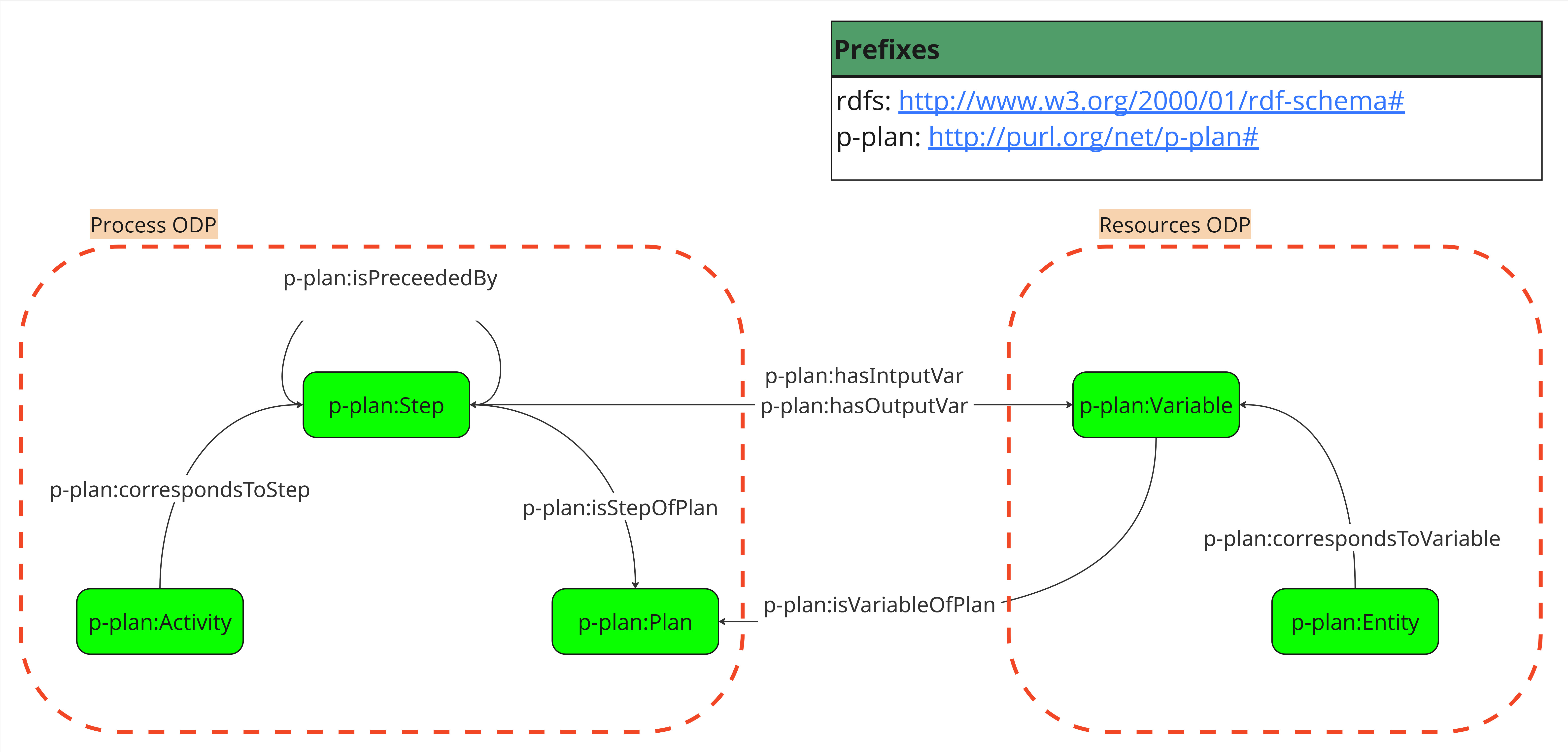}
    \caption{Extracted Process and Resource ODPs from P-PLAN based on defined requirements.}
    \label{fig:P-plan_patterns}
\end{figure}

\paragraph{Metadata4Ing (M4I)}  
M4I provides strong support for both the Resource and Project ODPs. It satisfies the Resource ODP requirements by linking experimental tools, methods, and variables via \texttt{m4i:hasEmployedTool}, \texttt{m4i:hasParameter}, and related properties. For Project ODPs it models project roles and participants using \texttt{schema:Project}, \texttt{prov:Association}, and \texttt{prov:Role}.
\begin{figure}[h]
    \centering
    \includegraphics[width=0.95\textwidth]{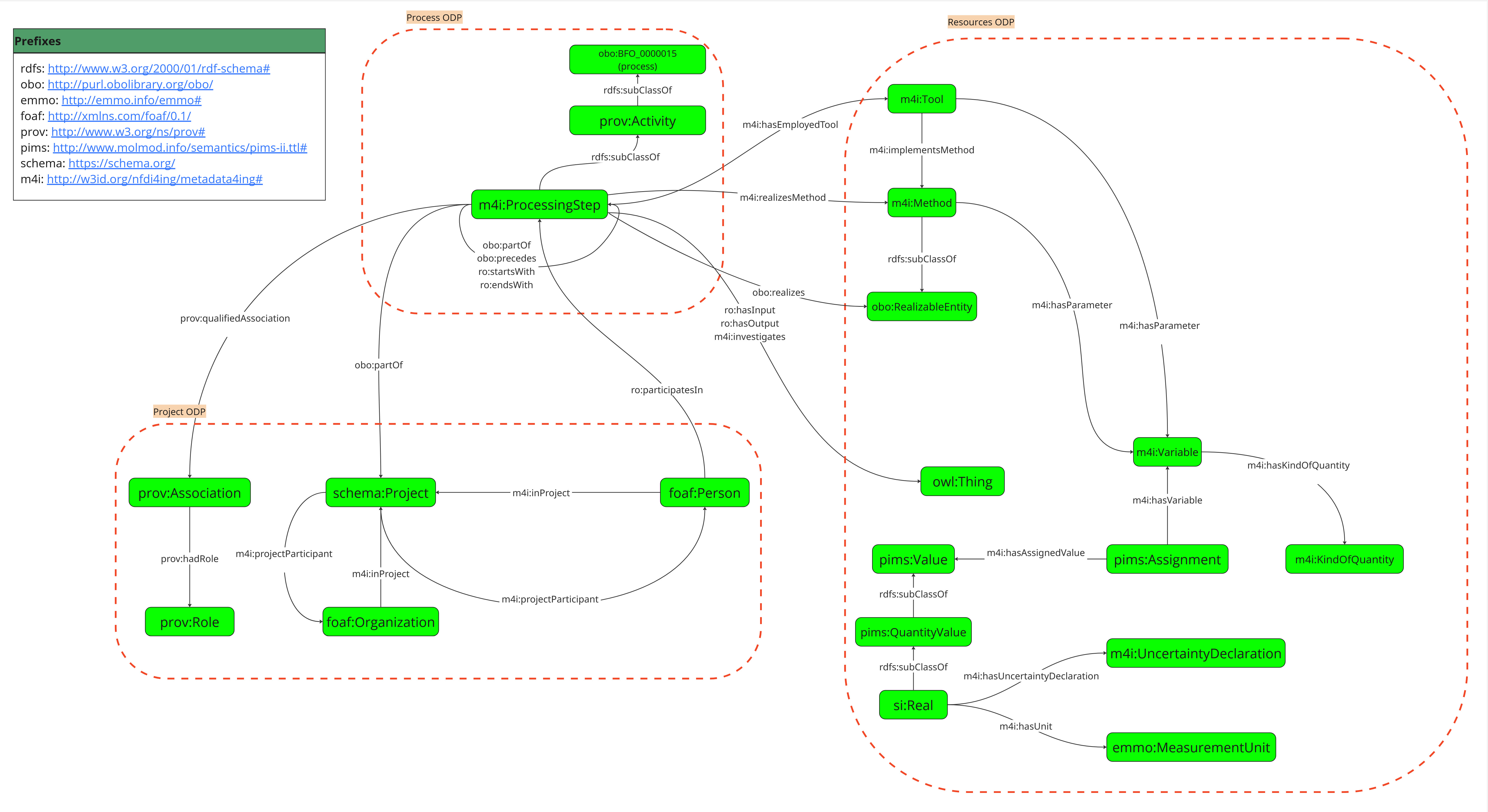}
    \caption{Extracted Resource and Project ODPs from Metadata4Ing.}
    \label{fig:Metadata4Ing_patterns}
\end{figure}

\paragraph{OPMW Ontology}  
OPMW addresses both the Process and Resource ODPs. It models the execution and templating of workflows using \texttt{opmw:WorkflowTemplateProcess}, \texttt{opmw:WorkflowExecutionProcess}, and their associated input/output artifacts. However, while agents are present, the model lacks detailed project-level constructs and necessary to fulfill Project ODP requirements.
\begin{figure}[h]
    \centering
    \includegraphics[width=0.9\textwidth]{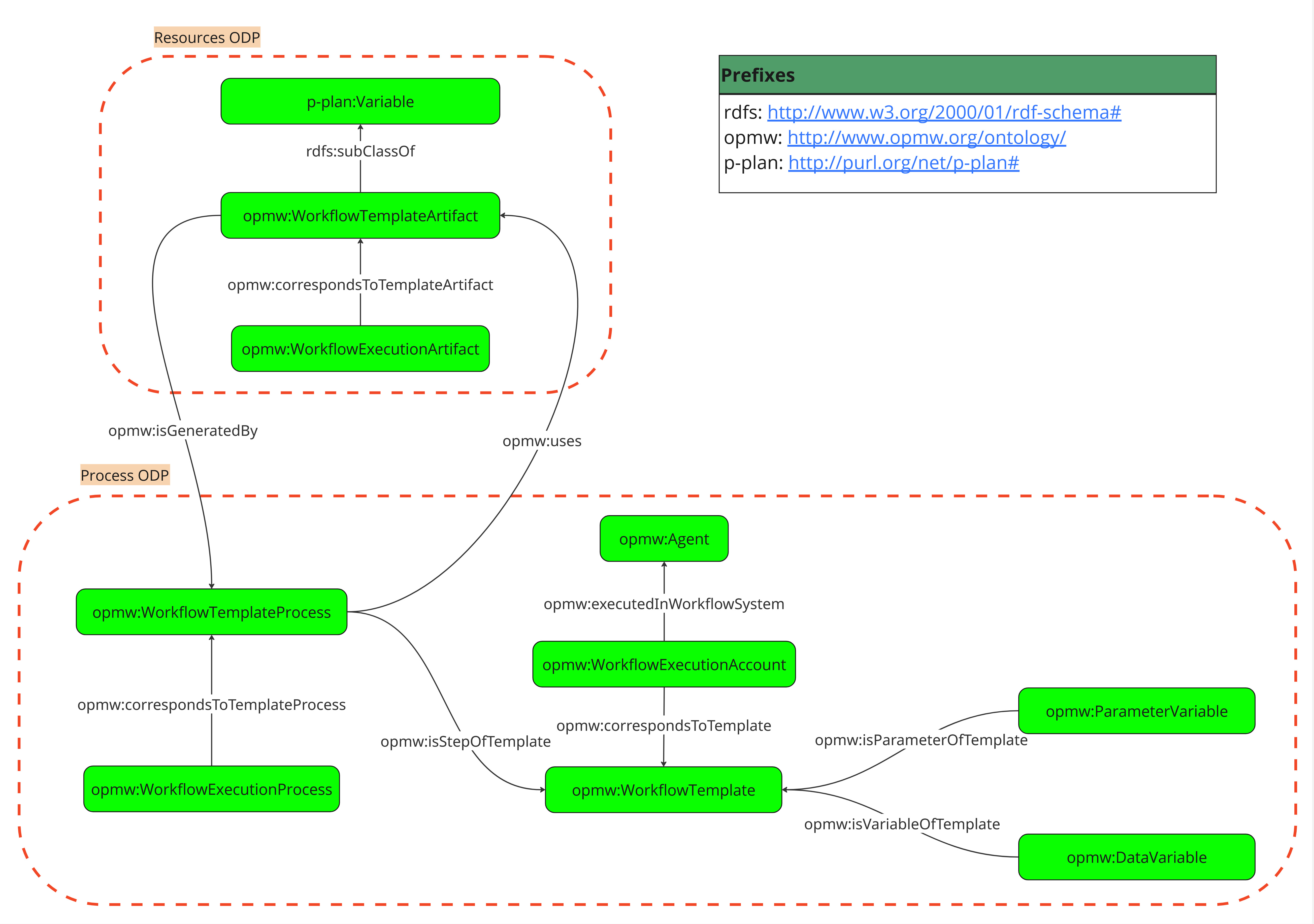}
    \caption{Extracted Process and Resource ODPs from OPMW.}
    \label{fig:opmw_patterns}
\end{figure}

\paragraph{Workflows in Linked Data (WILD)}  
WILD models workflow activities using classes like \texttt{wild:Activity}, \texttt{wild:WorkflowModel}, and their subclasses (e.g., \texttt{wild:CompositeActivity}, \texttt{wild:SequentialActivity}). However, despite this structural expressiveness, the ontology lacks alignment with concrete input/output or project-level constructs. Consequently, while it supports Process ODP, it does not sufficiently fulfill Resource and Project ODPs.
\begin{figure}[h]
    \centering
    \includegraphics[width=0.95\textwidth]{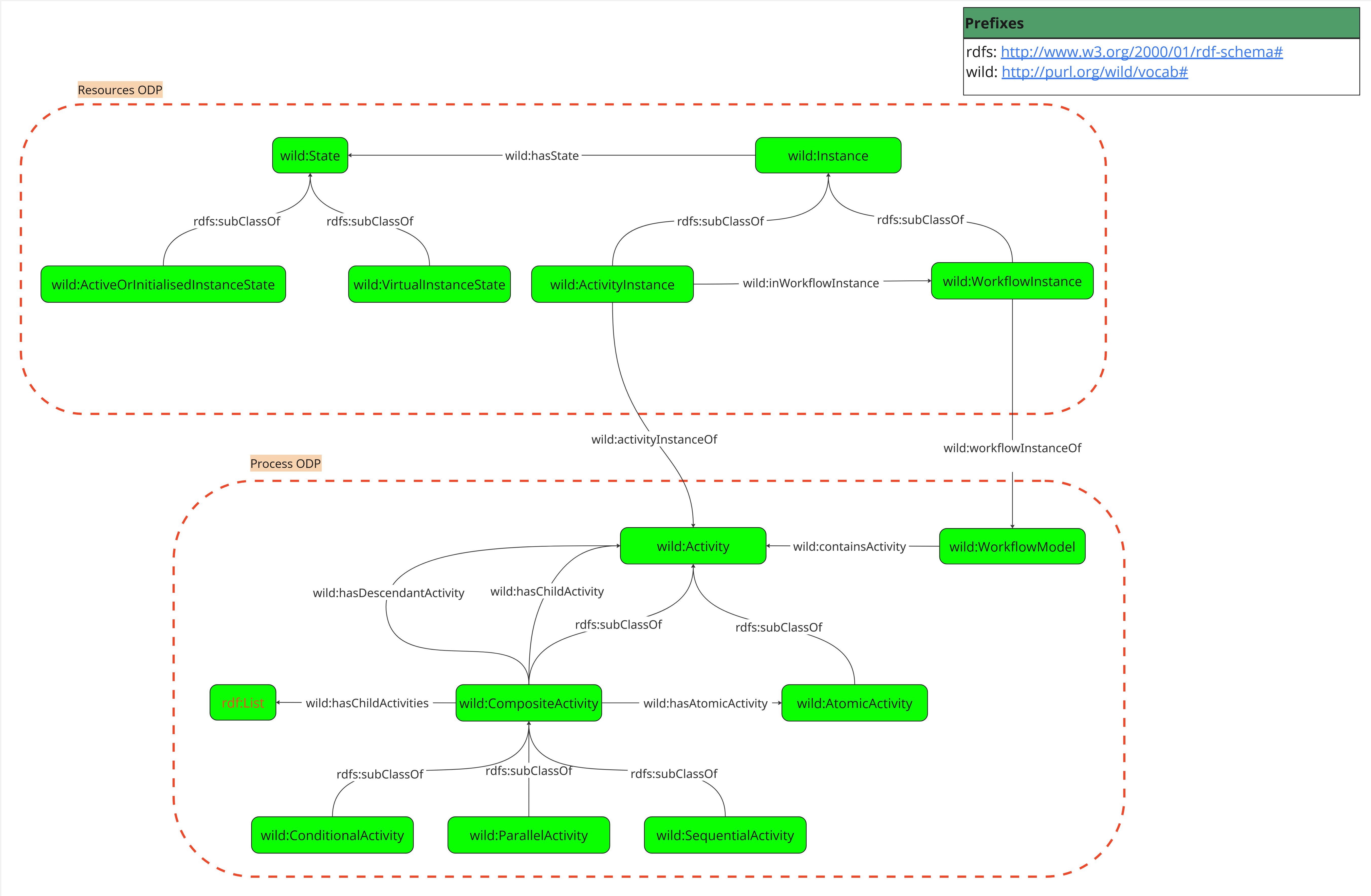}
    \caption{Extracted Process ODPs from WILD ontology.}
    \label{fig:wild_patterns}
\end{figure}

\paragraph{General Process Ontology (GPO)}  
GPO supports both the Process and Resource ODPs. The \texttt{gpo:Process} class and its related elements (\texttt{gpo:SubProcess}, \texttt{gpo:hasPredecessor}, \texttt{gpo:hasProcessInput}) provide a robust structure for process execution. While inputs, outputs, and tools are also modeled, the ontology lacks constructs for agent participation or organizational context and therefore does not fulfill the Project ODP requirements.
\begin{figure}[h]
    \centering
    \includegraphics[width=0.9\textwidth]{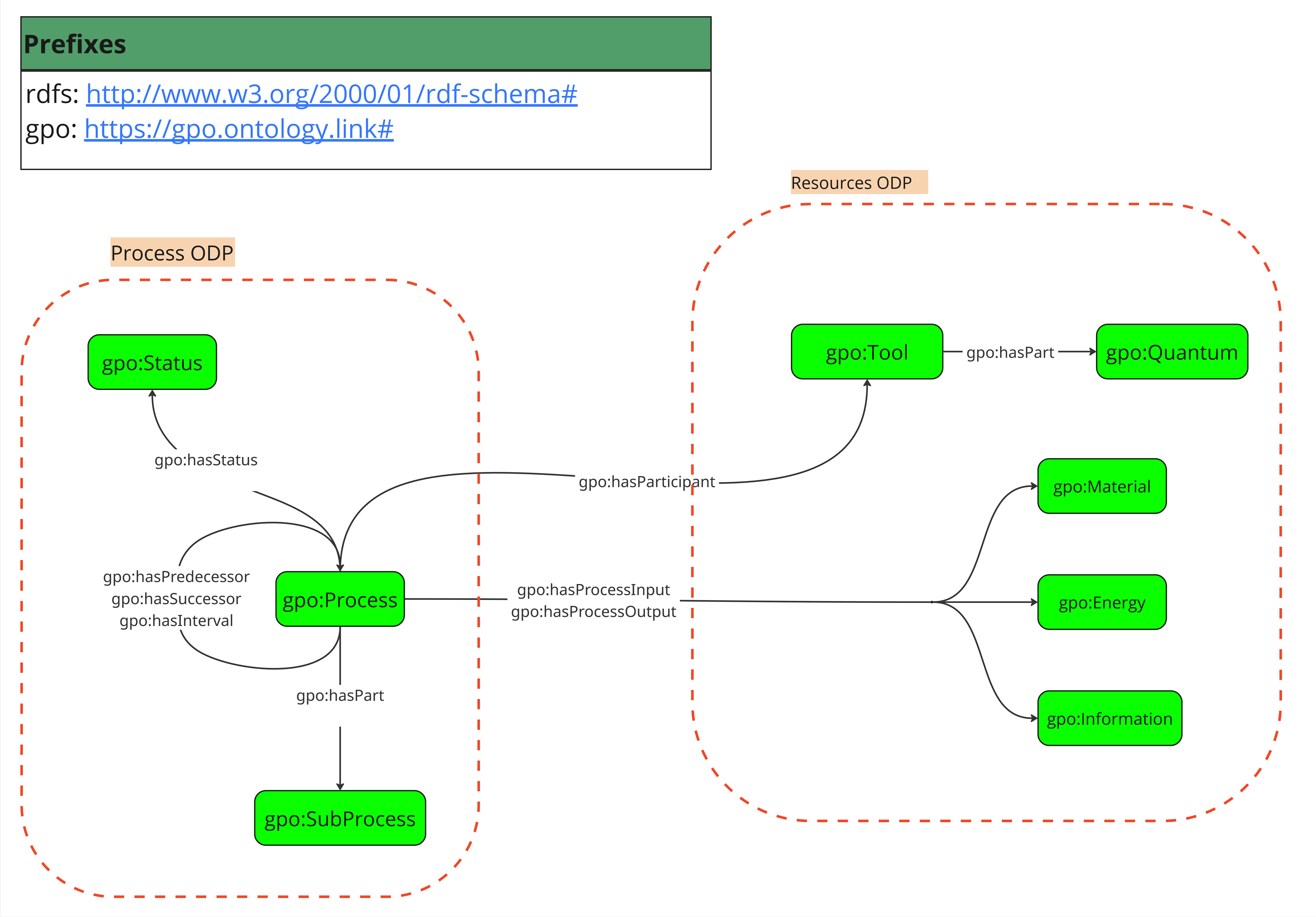}
    \caption{Extracted Process and Resource ODPs from GPO.}
    \label{fig:GPO_patterns}
\end{figure}

\paragraph{PMDcore Ontology}  
PMDcore supports all three ODP categories. It models process breakdown using \texttt{pmdco:Process} and \texttt{pmdco:ProcessingNode}, while \texttt{pmdco:ValueObject}, \texttt{pmdco:Object}, and their associated roles fulfill Resource ODP requirements. To support the Project ODP requirements, \texttt{pmdco:Project} and \texttt{pmdco:ProjectIdentifier} represent project membership and identifiers, enabling linkage of processes to organizational contexts.
\begin{figure}[h]
    \centering
    \includegraphics[width=0.95\textwidth]{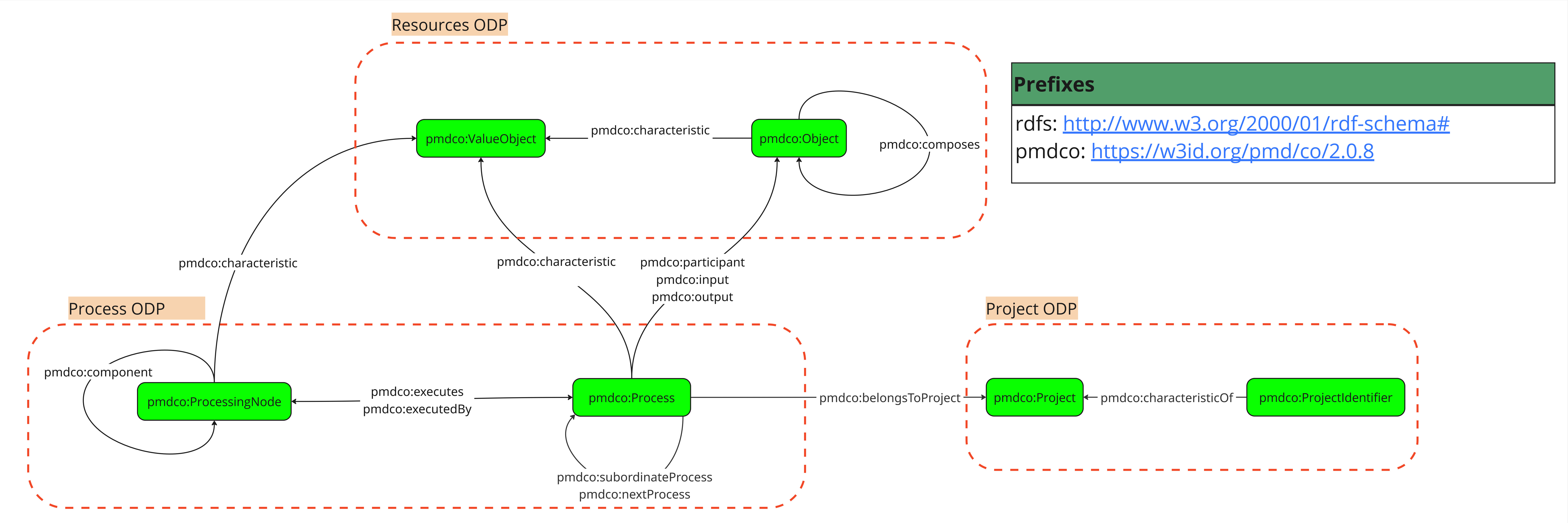}
    \caption{Extracted Process, Resource, and Project ODPs from PMDcore.}
    \label{fig:PMD_patterns}
\end{figure}

\subsection{Evaluation Results}
\label{sec:results_evaluation}

To assess how well each ontology supports the defined process modeling requirements in MSE, we applied a semantic similarity-based matching approach. For each requirement, we curated a set of ground truth IRIs (GT) and then retrieved a ranked list of relevant IRIs (Ext) using the \texttt{hkunlp/instructor-large}\cite{su2022one_instruct} embedding model, as introduced in Section~\ref{sec:baseline}. It is important to note that our evaluation considers all IRIs, including both classes and object properties.

We evaluated the performance of this extraction using precision (P), recall (R), and F$_1$-score ($F_1$) by comparing the extracted IRIs against the ground truth. A dash (–) indicates that an ontology lacks adequate coverage for the given requirement, meaning no ODP could be extracted in that category. Table~\ref{tab:ontology-eval} summarizes these results for all evaluated ontologies and requirement groups.

\begin{table}[ht]
\centering
\caption{Evaluation of ODP extraction across requirement categories using \texttt{hkunlp/instructor-large}. P = Precision, R = Recall, $F_1$ = F$_1$-score, GT = Number of ground truth IRIs, Ext = Number of extracted IRIs. Many Ext values equal 20, which corresponds to the fixed retrieval size in our semantic similarity pipeline. This cutoff was chosen to balance coverage and noise, ensuring each requirement is matched against a manageable subset of ontology terms.}
\label{tab:ontology-eval}
\resizebox{\textwidth}{!}{
\begin{tabular}{lccccc|ccccc|ccccc|}
\toprule
\textbf{Ontology} & \multicolumn{5}{c}{Process ODP (Req. 1)} & \multicolumn{5}{c}{Resource ODP (Req. 2)} & \multicolumn{5}{c}{Project ODP (Req. 3)} \\
 & P & R & $F_1$ & GT & Ext & P & R & $F_1$ & GT & Ext & P & R & $F_1$ & GT & Ext \\
\midrule
GPO & 0.10 & 0.17 & 0.12 & 12.0 & 20.0 & 0.05 & 0.11 & 0.07 & 9.0 & 20.0 & – & – & – & – & – \\
P-Plan & 0.56 & 0.62 & 0.59 & 8.0 & 9.0 & 0.44 & 0.67 & 0.53 & 6.0 & 9.0 & – & – & – & – & – \\
WILD & 0.00 & 0.00 & 0.00 & 11.0 & 0.0 & 0.00 & 0.00 & 0.00 & 9.0 & 0.0 & – & – & – & – & – \\
M4I & 0.05 & 0.12 & 0.07 & 8.0 & 20.0 & 0.25 & 0.28 & 0.26 & 18.0 & 20.0 & 0.15 & 0.25 & 0.19 & 12.0 & 20.0 \\
OPMW & 0.45 & 0.69 & 0.55 & 13.0 & 20.0 & 0.15 & 0.60 & 0.24 & 5.0 & 20.0 & – & – & – & – & – \\
PMDcore & 0.25 & 0.62 & 0.36 & 8.0 & 20.0 & 0.15 & 0.33 & 0.21 & 9.0 & 20.0 & 0.15 & 0.75 & 0.25 & 4.0 & 20.0 \\
\bottomrule
\end{tabular}
}
\end{table}

Our baseline results confirm that semantic similarity techniques can be applied to support the extraction of ontology design patterns aligned with defined requirements. In terms of F$_1$-score, P-PLAN achieved the top scores for the Process and Resource ODPs, demonstrating high precision. PMDcore supported all three ODPs (Process 0.36; Resource 0.21; Project 0.25) and showed strong recall for the Project ODP. M4I performed moderately for the Resource and Project ODPs but showed limited performance on the Process ODP.

However, the method has several limitations. First, ontologies with more descriptive labels and definitions tend to yield higher scores, introducing a bias based on annotation quality. Second, the formulation of requirements can influence the results by favoring certain ontologies that use specific terminology. To address these sources of bias in future work, several strategies can be pursued. First, the reliance on descriptive labels and definitions could be mitigated by incorporating structural features of the ontology (e.g., graph topology, axiomatic richness) alongside textual embeddings. Second, diversifying the set of requirement formulations by paraphrasing or using controlled vocabularies can reduce the sensitivity of results to specific wording choices. Finally, cross-ontology validation, where extraction results are compared against multiple ground truths curated by different expert groups, would help balance annotation subjectivity and improve the robustness 
of the evaluation.

A further limitation is that our current workflow processes each ontology independently, which makes it difficult to identify patterns that are shared or distributed across multiple sources. In addition, our focus on ontology modularization as the primary means of extracting reusable ODPs narrows the scope of reusability strategies. While modularization facilitates compact module creation and automated processing, it is not the only pathway to reuse. Alternative approaches include the direct publication of curated pattern catalogs, the use of competency questions to guide pattern identification, and the incorporation of ontology mappings to capture cross-ontology patterns. Despite these limitations, the proposed method shows promise as a first step toward 
automating the identification of reusable semantic structures for process modeling in MSE.

\section{Conclusion and Future Work}
\label{sec:Conclusion}

This work surveyed and evaluated ontologies and ontology design patterns (ODPs) relevant to process modeling in Materials Science and Engineering (MSE). We defined a set of domain-specific requirements to guide the assessment of existing ontologies and to support the identification of reusable semantic patterns. Based on this framework, process-related ODPs—covering aspects such as process structure, resource involvement, and project context—were extracted from relevant ontologies and consolidated into ground truth datasets.

Building on this manual analysis, we introduced and evaluated a baseline method for the automatic extraction of ODPs. This approach leverages semantic similarity techniques and modular ontology extraction to support scalable pattern identification aligned with domain requirements.

Future work will expand this analysis to additional MSE subdomains and modeling needs, such as value specification and property mapping. Beyond expanding the corpus, we plan to improve the methodology itself by integrating ontology mappings into the extraction workflow, allowing shared patterns to be identified across multiple ontologies rather than in isolation. We also intend to enhance the evaluation procedure by addressing the influence of annotation bias and requirement formulation, for example, through controlled vocabularies and multi-expert validation. In parallel, a platform will be developed to publish the extracted patterns, provide usage scenarios, and enable community feedback, thereby improving both methodological rigor and practical accessibility for the broader MSE community.

\begin{acknowledgments}
  This publication was written by the NFDI-MatWerk consortium as being part of the German National Research Data Infrastructure (NFDI) programme. NFDI is financed by the Federal Republic of Germany and the 16 federal states and funded by the Federal Ministry of Education and Research (BMBF) – funding code M532701 / the Deutsche Forschungsgemeinschaft (DFG, German Research Foundation) - project number 460247524. The authors would like to thank the participants of the Linked Open Data Working Group (LOD-WG) within NFDI-MatWerk for their valuable feedback and discussion during the evaluation of the ontology design patterns.
\end{acknowledgments}


\section*{Declaration on Generative AI}
 \noindent During the preparation of this work, the author(s) used GPT-4 and Claude Sonnet~4 in order to: Grammar and spelling check. After using these tool(s)/service(s), the author(s) reviewed and edited the content as needed and take(s) full responsibility for the publication’s content. 

\bibliography{sample-ceur}



\end{document}